\documentclass[letterpaper,oneside,11pt,pdftex]{article}

\usepackage[english]{babel}
\usepackage{textcomp}
\usepackage{gensymb}
\usepackage[justification=justified,margin=1cm,font=small,labelfont=bf,textfont=rm]{caption}

\usepackage[letterpaper,top=2cm,bottom=2cm,left=2cm,right=2cm,marginparwidth=1.75cm]{geometry}

\usepackage{amsmath}
\usepackage{graphicx}
\usepackage[pdftex,bookmarksnumbered=true,breaklinks=true,colorlinks=true, allcolors=blue]{hyperref}
\usepackage{array}
\usepackage{multirow}
\providecommand{\keywords}[1]
{\begin{center}
  \vspace{-0.1cm}
  \small
  \textbf{\textit{Keywords---}} #1
  \vspace{-0.2cm}
\end{center}}
\usepackage[numbers,square,comma,sort&compress]{natbib}
\usepackage{IEEEtrantools}
\makeatletter
\AtBeginDocument{
  \hypersetup{
    pdftitle = {\@title},
    pdfauthor = {\@author}
  }
}
\makeatother

\title{\texorpdfstring{\vspace{-1em}\begin{flushright}
			{\small LA-UR-24-29632}
		\end{flushright}\vspace{1em}}{}%
Exploring the relation between transonic dislocation glide and stacking fault width in FCC metals}
\author{Kathryn R. Jones, Khanh Dang, Daniel N. Blaschke, Saryu J. Fensin, Abigail Hunter}
\date{December 6, 2024}

\graphicspath{{./}{./figures/}}
\begin{document}
%%% only use if \bibliographystyle{IEEEtran} below
\bstctlcite{IEEEexample:BSTcontrol}
%%%
\newcolumntype{F}[1]{%
    >{\raggedright\arraybackslash\hspace{0pt}}p{#1}}%
\newcolumntype{T}[1]{%
    >{\centering\arraybackslash\hspace{0pt}}p{#1}}%
\maketitle

\begin{center}
	\renewcommand{\thefootnote}{\fnsymbol{footnote}}
	\vspace{-0.3cm}
	Los Alamos National Laboratory, Los Alamos, NM, 87545, USA
	\\[0.5cm]
	\ttfamily{E-mail: krjones2@uw.edu, kqdang@lanl.gov, dblaschke@lanl.gov, saryuj@lanl.gov, ahunter@lanl.gov}
\end{center}
%\doublespacing
%\linenumbers
\begin{abstract}
Theory predicts limiting gliding velocities that dislocations cannot overcome. Computational and recent experiments have shown that these limiting velocities are soft barriers and dislocations can reach transonic speeds in high rate plastic deformation scenarios. In this paper we systematically examine the mobility of edge and screw dislocations in several face centered cubic (FCC) metals (Al, Au, Pt, and Ni) in the extreme large-applied-stress regime using MD simulations. Our results show that edge dislocations are more likely to move at transonic velocities due to their high mobility and lower limiting velocity than screw dislocations. Importantly, among the considered FCC metals, the dislocation core structure determines the dislocation's ability to reach transonic velocities. This is likely due to the variation in stacking fault width (SFW) due to relativistic effects near the limiting velocities.
\end{abstract}

\keywords{transonic dislocation glide, stacking fault width, molecular dynamics, FCC metals}

\hspace{5em}
\tableofcontents
\pagebreak

\section{Introduction}

Since its first discovery by Taylor in 1934 \cite{Taylor:1934}, dislocation  properties have been extensively studied due to their important role in plasticity. However, there are still unresolved questions about dislocation behaviors, especially under extreme loading conditions such as high applied strain rates \cite{Gurrutxaga:2020,Pellegrini:2023}.
In this regime, one of the key questions is whether dislocations can reach transonic and/or supersonic speeds under sufficiently high applied stress. Some early experiments suggested that there may be a limiting velocity of dislocation movement \cite{Parameswaran1971,Weertman1973}. However, due to the inherent challenges with tracking dislocations in this extreme regime, the first experimental observation of transonic dislocations only happened recently in diamond \cite{katagiri2023transonic}. Therefore, a majority of the studies of dislocation behaviors in this regime have relied on theoretical and modeling tools such as molecular dynamics (MD) simulations.

Theoretically, the existence of transonic and supersonic dislocations, which are separated from the subsonic ones by critical velocity ``barriers'',  was predicted by the steady-state theory of perfect dislocations \cite{weertman1981moving,Rosakis200195}.
Importantly, these velocity barriers have been speculated to be soft limits that can be overcome under extreme conditions \cite{MARKENSCOFF20082225,Blaschke2021}. This idea was supported by results from MD simulations for 
FCC metals such as Ni, Al, and Cu \cite{Groh_2009,Olmsted_2005, DAPHALAPURKAR2014, Marian2006, TSUZUKI2009, Peng2019, Blaschke2021, Oren_2016,Young2004}, BCC W \cite{Gumbsch1999,Li2002}, and HCP Mg \cite{DANG2022}. A majority of transonic dislocations are only observed in edge dislocations where the subsonic dislocations asymptotically approach the lowest limiting velocities until some critical stress above which transonic motion becomes possible \cite{Groh_2009,Marian2006, TSUZUKI2009, Peng2019, Blaschke2021, Oren_2016,Young2004,DANG2022}. On the other hand, for screw dislocations, instabilities in the form of additional dislocation nucleation from the existing partials or homogeneously within the simulation box appeared at higher applied shear stresses ($>$500 MPa) in both FCC and HCP metals \cite{ Blaschke2021, Oren_2016,Olmsted_2005,DANG2022}. In fact, transonic speeds were only observed in Cu screw dislocations at very low temperatures ($<$10K) \cite{Marian2006, TSUZUKI2009, Peng2019, Blaschke2021, Oren_2016,Olmsted_2005}.  
It was suggested that at high temperature, the screw dislocations experience larger phonon drag, which prevents them from overcoming the limiting velocities before instabilities occur \cite{ Blaschke2021}.
However, it remains unclear which parameters govern the ability of one material system's dislocation to reach transonic gliding speeds.

One of the challenges is the lack of a systematic study across different material systems. Dislocation motion via MD simulations can be studied using different simulation setups \cite{Marian2006, TSUZUKI2009, Peng2019,  Oren_2016,Olmsted_2005,Bertin:2020}. For instance, there are several methods to apply a shear driving force to move the dislocation such as (i) adding a shear force to the top layer of atoms while fixing the bottom layer of atoms, (ii) applying a shear deformation to the simulation cell, and (iii) applying a velocity to the top layer of atoms to impose a shear strain rate.
Moreover, the boundary conditions between fixed versus free boundaries also seem to affect the dislocation mobility at near-sonic velocities \cite{DUONG2023105422,DUONG2024120050,Blaschke:2023rad}.
Importantly, while there have been MD studies that investigate the behavior of dislocations in the transonic regime \cite{Groh_2009,Marian2006, TSUZUKI2009, Peng2019, Blaschke2021, Oren_2016,Young2004,DANG2022}, they used different interatomic potentials and material systems which prevents a comprehensive understanding of which parameter affects the ability of dislocations to accelerate beyond the subsonic regime. 

Therefore, the goal of this paper is to explore the governing parameters that control if the dislocation can reach transonic velocities. To achieve this, mobilities of dislocations under high applied shear stresses are systematically investigated in FCC metals such as Al, Au, Pt, and Ni. These material systems and the corresponding interatomic potentials are carefully chosen so that the important dislocation properties such as unstable stacking fault (USF) energy, intrinsic stacking fault (ISF) energy, and shear modulus are varied.

\section{Methodology: Molecular Dynamics Simulations}

MD simulations are performed using the classical molecular dynamics simulation code LAMMPS \cite{LAMMPS} and visualized using the Open Visualization Tool (OVITO) \cite{ovito}.
Table \ref{table:1} shows the list of interatomic potentials used in this work with important material properties for dislocations such as (average) shear modulus, ISF energy, USF energy, and equilibrium stacking fault widths (SFW) (which were determined for edge dislocations at 0K).
Only FCC metals with the same functional form for the interatomic potentials are used to reduce the complexity of the comparison.
For Pt, due to the lack of reasonable embedded atom model (EAM) potentials, a modified EAM (MEAM) is used instead. Importantly, the four FCC material systems and corresponding interatomic potentials are specifically chosen to have a wide range of values for shear modulus, ISF energy, USF energy, and equilibrium SFW.   

\begin{table}[!ht]
	\centering
	\caption{Potentials of metals used in this work. Three elastic constants were used from the potential file as well as the intrinsic stacking fault (ISF) energy; the shear modulus $G$ was determined by the Kr\"oner average for cubic crystals, see \cite{Blaschke:2017Poly}.
    Density and limiting velocity were calculated using the data from the potential file. The equilibrium stacking fault width (SFW) was found by running the edge dislocation simulations at a shear stress of 0MPa and 0K.}
	\label{table:1}
	\small
	\begin{tabular}{|F{0.07\textwidth}|T{0.09\textwidth}|T{0.06\textwidth}|T{0.06\textwidth}|T{0.06\textwidth}|T{0.06\textwidth}|T{0.08\textwidth}|T{0.09\textwidth}|T{0.08\textwidth}|T{0.09\textwidth}|}
		\hline
		Metal [Potential]& 
		Lattice Constant (\AA)& 
		$C_{11}$ (GPa)&
		$C_{12}$ (GPa)& 
		$C_{44}$ (GPa)& 
            $G$ (GPa) &
		Intrinsic SFE (mJ/m\textsuperscript{2})&
            Unstable SFE (mJ/m\textsuperscript{2})&
		Density (g/cm\textsuperscript{3})&
		Equilbrium SFW (\AA)\\ [0.5ex]
		\hline
		Ni \cite{Ni_Potential} & 3.52 & 247.86 & 147.83 & 124.84 & 87.2 & 125.25 & 367.8 & 8.94 & 31.37 \\ 
		\hline
		Pt \cite{Pt_Potential} & 3.92 & 358.12 & 253.51 & 77.55 & 66.4 &110 & 596.0 \cite{WEI20071489} & 21.55 & 16.20\\
		\hline
		Au \cite{Au} & 4.07 & 201.65 & 169.53 & 45.97 & 30.8 & 42.58 & 91.69 & 19.4 & 67.04 \\ [1ex]
		\hline
		Al \cite{AlPotential.68.024102} & 4.05 & 116.82 & 60.12 & 31.66 & 30.3 & 113.77 & 149.72 & 2.70 & 22.55\\ [1ex]
		\hline
	\end{tabular}
\end{table}

\begin{table}[!ht]
	\centering
	\caption{Crystal lattice orientation in the $X$, $Y$ and $Z$ direction for edge and screw dislocations as well as the limiting velocities for each metal and type of dislocation, calculated using \cite{pydislocdyn}, see also \cite{Blaschke_2021}.}
	\label{table:2}
	\small
	\begin{tabular}
		{|F{0.07\textwidth}|T{0.07\textwidth}|T{0.05\textwidth}|T{0.05\textwidth}|T{0.05\textwidth}|T{0.15\textwidth}|T{0.2\textwidth}|}
		\hline
		\multicolumn{1}{|c|}{Metal} & \multicolumn{1}{c|}{Dislocation Type} & \multicolumn{1}{c|}{$X$} & \multicolumn{1}{c|}{$Y$} & \multicolumn{1}{c|}{$Z$} & 1st Limiting Velocity (km/s) & \multicolumn{1}{c|}{2nd Limiting Velocity (km/s)} \\ \hline
		\multirow{2}{*}{Ni}           &   Edge    &  [1$\overline{2}$1]   &    [111]    &   [$\overline{1}$01]   &  2.37  &  3.74 \\ \cline{2-7} 
		&     Screw   &[$\overline{1}$01]  &   [111]   &   [$\overline{1}$2$\overline{1}$]  &    3.05  & N/A    \\ \hline
		\multirow{2}{*}{Pt}           &   Edge    &  [1$\overline{2}$1]   &    [111]    &   [$\overline{1}$01]   &   1.55 & 1.90    \\ \cline{2-7} 
		&    Screw   & [$\overline{1}$01]   &  [111]   &   [$\overline{1}$2$\overline{1}$]  &   1.76  & N/A    \\ \hline
		
		\multirow{2}{*}{Au}           &   Edge    &  [1$\overline{2}$1]   &    [111]    &   [$\overline{1}$01]   &   0.91   &  1.54  \\ \cline{2-7} 
		&    Screw   & [$\overline{1}$01]   &  [111]   &   [$\overline{1}$2$\overline{1}$]  &   1.21  & N/A     \\ \hline
		
		\multirow{2}{*}{Al}           &   Edge    &  [1$\overline{2}$1]   &    [111]    &   [$\overline{1}$01]   &   3.24  &  3.42   \\ \cline{2-7} 
		&    Screw   & [$\overline{1}$01]   &  [111]   &   [$\overline{1}$2$\overline{1}$]  &   3.36  & N/A     \\ \hline
	\end{tabular}
\end{table}

\begin{figure}[ht]
\centering
\includegraphics[width=0.6\linewidth]{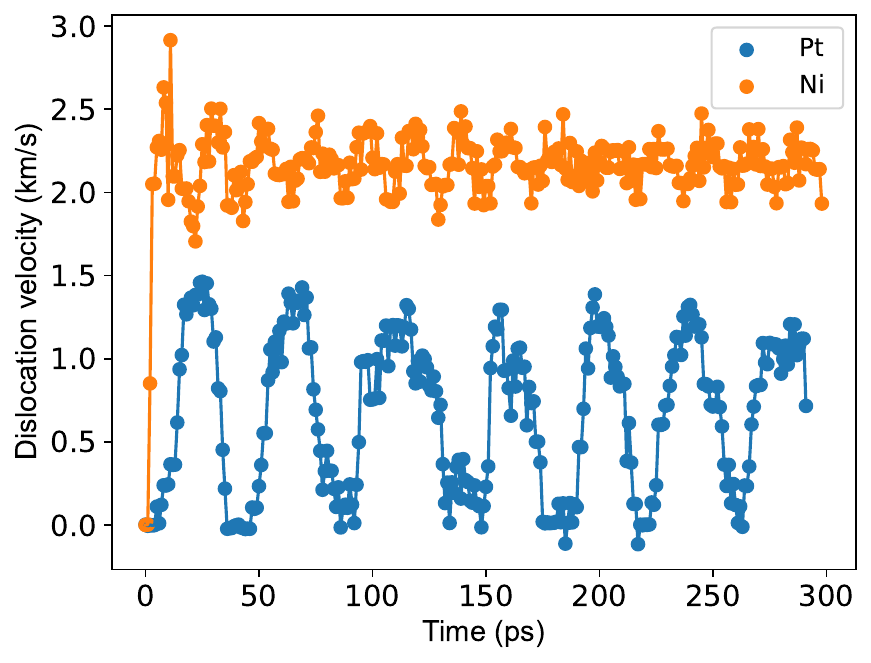}
\caption{We highlight the difference in edge dislocation motion between Ni, where dislocation glide quickly converges to an (almost) steady state, and Pt, which exhibits a ``stick-slip'' behavior.}
\label{fig:stickslip}
\end{figure}

To model the dislocation mobility, simulation cells with specific orientations are used for edge and screw dislocations as shown in Table \ref{table:2}. Table \ref{table:2} also includes the first two limiting velocities for the edge and screw dislocations.
The lowest limiting velocities for these slip systems can be determined analytically by \cite{Blaschke_2021,Blaschke2021}
\begin{align}
v_\text{lim}^\text{FCC,edge}&=\sqrt{\frac{\min(C_{44},C')}{\rho}}
\,,&
v_\text{lim}^\text{FCC,screw} &=\sqrt{\frac{3C'C_{44}}{\rho(C_{44}+2C')}}
\,,
\end{align}
where $\rho$ denotes the material density and $C'=(C_{11}-C_{12})/2$.
The higher limiting velocities were determined numerically as detailed in the review article \cite{Blaschke_2021} using the Python code PyDislocDyn \cite{pydislocdyn}.
Note that all metals considered in this work have a Zener %\footnote{This ratio, which in the modern literature is named after Clarence Zener (who introduced the symbol $A$ in his 1948 book) \cite{Zener:1948,Zener:1947}, was in fact first used as a measure of anisotropy for cubic crystals by Fuchs \cite{Fuchs:1936,Blackman:1955} several years earlier.}
anisotropy ratio $A=C_{44}/C'>1$ so that the lowest limiting velocity for edge dislocations depends on $C'$.

The simulation box is approximately 16 x 52 x 550 \AA{} along the $X$, $Y$, and $Z$ directions, respectively. Periodic boundary conditions were applied along the  $X$ and $Z$ directions so the dislocation can glide in and out of the simulation box without encountering barriers. Fixed boundary conditions are applied along the $Y$ direction with appropriate pre-strain to alleviate stress caused from thermal expansion. A single dislocation is inserted at the center of the simulation box using the Volterra displacement field with appropriate boundary treatment as detailed in Refs. \cite{dang2018pressure,DANG2019}. Shear forces are applied to the top layer atoms to move the dislocations. Once the dislocation motion reaches equilibrium, the dislocation position and SFWs are tracked and determined by using the common neighbor analysis in LAMMPS \cite{LAMMPS} to identify perfect and defected atoms.
While this method is efficient, as temperature and stress increased there were some areas where the lattice shifted enough due to lattice vibrations that it was no longer considered a FCC structure, which complicates the dislocation core region tracking.
Data points influenced by this are removed from the analysis.
Since the dislocation cores remain planar, only the SFW is used to describe the change in the dislocation core structure.
The average velocities and SFWs are recorded only after the dislocation reaches steady state (after 50 ps) for all cases except at 100K for both edge and screw dislocations in Pt where the full range of data (0 to 300 ps) is included. For these two cases, the dislocation motion transitions into stick-slip behavior instead of steady state glide as seen in Fig. \ref{fig:stickslip}.
Understandably, this will slightly overestimate the error for these two cases.  Importantly, the averaged velocities do not vary significantly regardless of ranges of recorded data. In order to keep the temperature constant, a Nos\'e-Hoover type thermostat and barostat were used \cite{Hoover} as described in Ref. \cite{DANG2019}. This allows slight changes in temperature while still keeping the temperature within a few degrees of the target. Temperature effects are considered by using two temperatures at 10 and 100K. The applied stresses are increased until the dislocation achieves transonic motion or the simulation exhibits instabilities.

\section{Results}

\subsection{Mobility of screw dislocations}

\begin{figure}[ht]
	\centering
	\includegraphics[width=\textwidth]{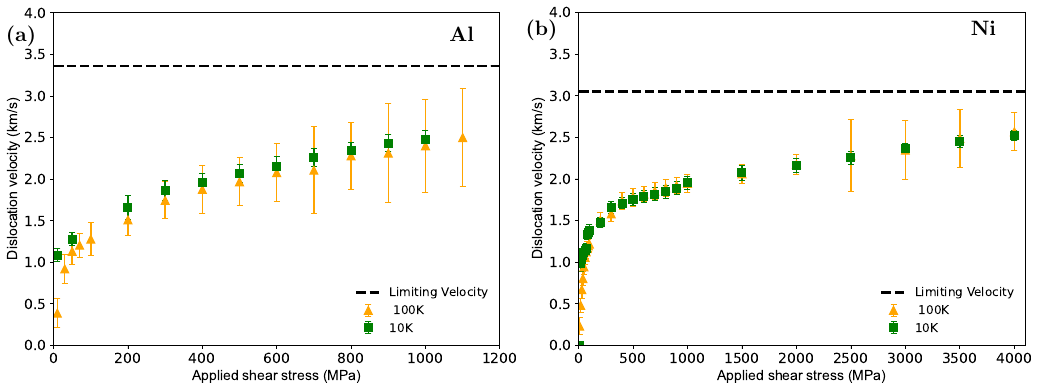}
	\includegraphics[width=\textwidth]{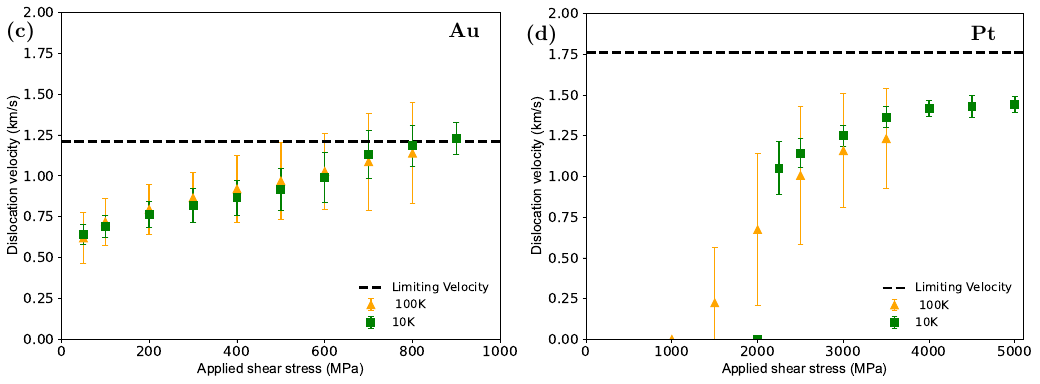}
	\caption{Velocity of screw dislocations in Al (a), Ni (b), Au (c) and Pt (d) vs. the shear stress applied. The dashed lines indicate the limiting velocities.}
	\label{fig_screw}
\end{figure}

Figure \ref{fig_screw} (a)--(d) shows the mobility curves for screw dislocations in Al, Ni, Au, and Pt.
Compared to Ni and Al, Au and Pt dislocations are less mobile. The lower mobilities of Au could be due to the relatively lower ISF energy and thus larger dislocation core structures compared to Al and Ni \cite{MILLS20012278}.
Among the four, Pt dislocations are extremely immobile with dislocation glide observed only for applied stress greater than 2.25 GPa at 10K and greater than 1.5 GPa at 100K.
The significantly higher Peierls stress of Pt dislocations could be due to the combination of high shear modulus (shown in Table \ref{table:1}) and small equilibrium SFW according to the analytical Peierls-Nabarro model.
For the same applied shear stress, the dislocation velocities at higher temperature (100K) also fluctuate more compared to the ones at 10K. 

For each material, the mobility curve contains two distinct regimes of phonon and radiative drag.
As the applied stress increases, the dislocation velocities start to plateau below the limiting velocities shown in Table \ref{table:2}. Higher applied shear stresses result in nucleation of additional dislocations or defects that prevent the proper tracking and determination of dislocation velocities, and thus these simulations are not included in the analysis.
Importantly, no transonic jump is observed for screw dislocations, though the highest screw dislocation velocity observed in Au at 10K barely exceeds the calculated limiting velocity (which is within the error bars).
For the other three FCC metals, no transonic dislocations are found.
This result is consistent with previous MD simulations for screw dislocations where only transonic dislocation motion is observed in Cu at very low temperature ($<$10K) \cite{Blaschke_2021}. 

\subsection{Mobility of edge dislocations}

\begin{figure}[ht]
	\centering
	\includegraphics[width=\textwidth]{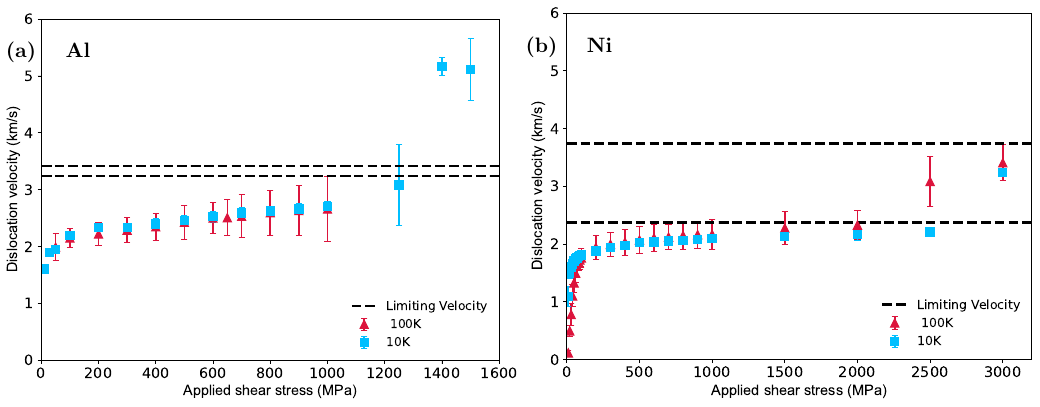}
	\includegraphics[width=\textwidth]{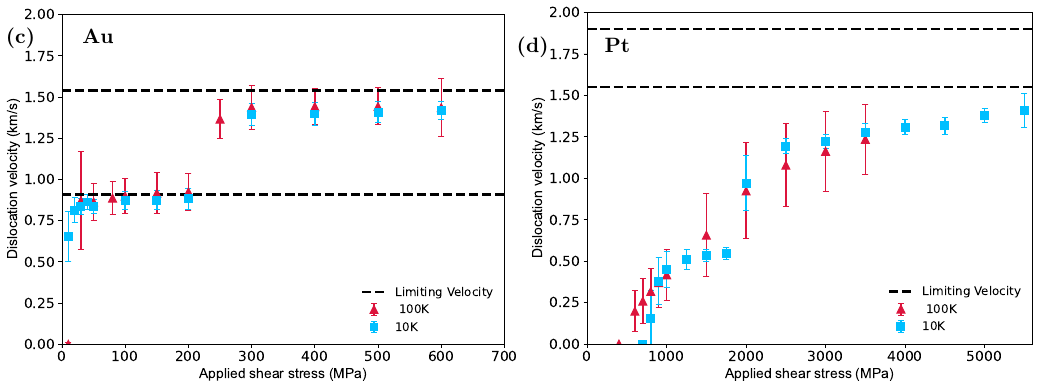}
	\caption{Velocity of edge dislocations in Al (a), Ni (b), Au (c) and Pt (d) vs. the shear stress applied. The dashed lines indicate the two lowest limiting velocities.}
	\label{fig_edge}
\end{figure}

\begin{figure}[ht]
	\centering
	\includegraphics[width=0.6\textwidth]{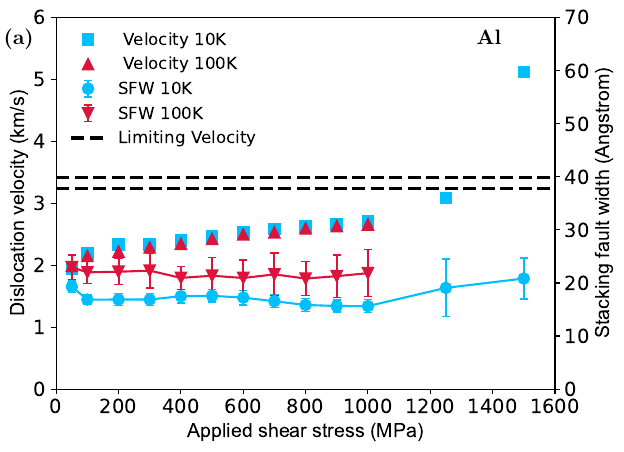}
	\includegraphics[width=0.6\textwidth]{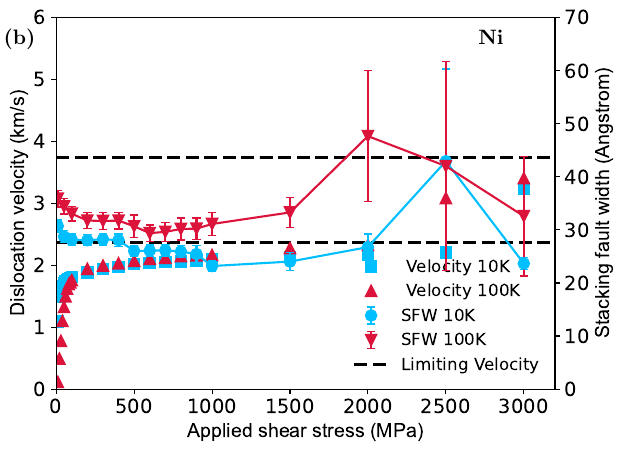}
	\caption{Velocity and stacking fault width of edge dislocations in Al (a) and Ni (b) vs. the shear stress applied. The dashed lines indicate the two lowest limiting velocities.}
	\label{fig_edge_Al_Ni}
\end{figure}

\noindent

Figure \ref{fig_edge} shows the mobility curves for edge dislocations in Al, Ni, Au, and Pt. Overall, edge dislocations for all considered FCC metals are more mobile than screw dislocations. Similar to screw dislocations, the mobility of edge dislocations also scale linearly with applied shear stress at low velocities and eventually plateau at or near the limiting velocities as the stress increases. However, unlike screw dislocations, transonic and supersonic velocities are observed in edge dislocations for large applied shear stresses.
For Al, there is a jump in the dislocation velocity to the (second) transonic regime at 10K with applied shear stress of 1.3 GPa, but there was no jump at 100K.
In contrast to Au and Ni discussed below, the first transonic regime is skipped, possibly due to to the first two limiting velocities being in close proximity of one another.
At 100K, additional dislocations start to form, glide, and interact with the existing edge dislocation.
This is different from previous MD simulations by Olmsted et al. \cite{Olmsted_2005} where the authors observed transonic speed of edge dislocations in Al at higher temperature ($>324$K).
This is due to the difference in the utilized interatomic potentials between the two studies.
As shown in Fig. \ref{fig_edge}, Au and Pt edge dislocations have a lower mobility compared to Al and Ni ones. Both reach velocities of approximately 1.4 km/s compared to 5 and 3.24 km/s for Al and Ni dislocations, respectively.
In particular, Pt dislocations are less mobile compared to the other three materials and only start gliding with applied shear stress above 600 and 300 MPa at 10 and 100K, respectively.
Importantly, even though the Pt edge dislocation is much more mobile compared to its screw counterpart, it also does not reach transonic velocity even at extremely high applied shear stress of 5 GPa.
Above this shear stress, instabilities occur via nucleation of additional dislocations. On the other hand, the Au edge dislocation does not only cross over the $1^\text{st}$ limiting velocity at applied shear stress greater than 200 MPa but also comes close to the $2^\text{nd}$ limiting velocity.

In order to understand the behavior of the dislocation core structure as the dislocations reach the limiting velocities, Fig. \ref{fig_edge_Al_Ni} (a)--(b) shows the mobility curves and SFWs of Al and Ni edge dislocations, respectively.
Initially, the SFW of the Al edge dislocations slightly reduce until the dislocation velocities reach approximately 2.7 km/s. At 1.25 GPa applied shear stress where the dislocation jumps to the transonic regime, the SFW slightly increases. For Ni, the dislocation velocity jumps to the transonic regime (at about 3.27 km/s) at an applied shear stress of 2 GPa and 2.5 GPa for temperatures of 10 and 100K, respectively, as shown in Fig. \ref{fig_edge_Al_Ni} (b). The transonic dislocation motion is consistent with previous MD results for edge dislocations in Ni using a different EAM potential \cite{Olmsted_2005}. The dislocation velocity jump happens at a lower stress for higher temperatures which could be due to the larger dislocation core as the temperature increases \cite{zhang2018temperature, LI2017}. Similar to Al, the Ni edge dislocation also reduces its SFW at first up to the velocity of about 2.09 km/s. From there, the SFW increases as the dislocation  gets closer to the limiting velocities. However, unlike Al, once the dislocation reaches transonic speed, its SFW contracts.
As they transition into the transonic regime, Ni edge dislocations also contract and extend more during their glide motions compared to the Al edge dislocations, which results in higher standard deviations in SFWs shown in Fig. \ref{fig_edge_Al_Ni}.

\begin{figure}[ht]
	\centering
	\includegraphics[width=0.6\textwidth]{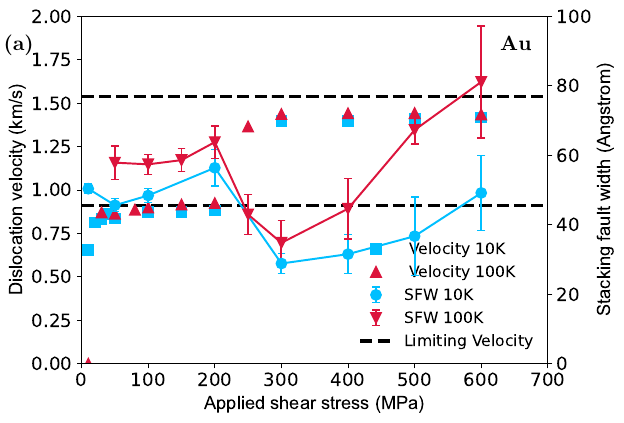}
	\includegraphics[width=0.6\textwidth]{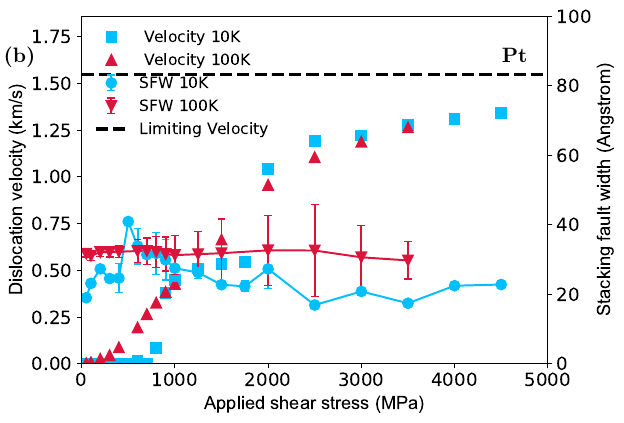}
	\caption{Velocity of edge dislocations in Au (a) and Pt (b) vs. the shear stress applied. The dashed lines indicate the (lowest) limiting velocities.}
	\label{fig_edge_Au_Pt}
\end{figure}

Similarly, Fig. \ref{fig_edge_Au_Pt} shows the mobility curves and SFWs of Au and Pt edge dislocations, respectively.
For Pt edge dislocations, the SFW at 100K remains relatively constant. At a lower temperature of 10K, the SFW increases up to an applied shear stress of 600 MPa, where the edge dislocation starts to move. Then, the SFW reduces back to the previous values and remains constant as the applied shear stress increases. On the other hand, the SFW of Au edge dislocations at both 10K and 100K reduces initially, then increases as the velocities approaches the $1^\text{st}$ limiting velocity of 0.91 km/s. As the dislocation velocities cross over this limiting velocity, the SFW reduces again until the applied shear stress approaches 300 MPa. For applied shear stresses greater than 300 MPa, the Au edge dislocation core starts to widen again as its velocity approaches the $2^\text{nd}$ limiting velocity of 1.54 km/s. The expansion and constriction of the SFW for moving dislocations at different velocities is likely due to the interplay between dislocation drag and relativistic effects \cite{relativistic1,Pellegrini:2018,Kim:2020}.

In closing, we note that the SFW of the screw dislocations look similar, i.e. slight initial contraction and subsequent hint of expansion (not shown), with the main exception that there is no contraction at very high stress, consistent with the absence of any transonic screw dislocation velocity.

%\clearpage
\section{Discussion}

\subsection{Variations in dislocation stacking fault width as the velocity becomes transonic}

The evolution of the SFWs seen in this study as the edge dislocations glide with subsonic to transonic velocities can be explained by the relativistic effects on dislocation core structures.
Specifically, the relativistic effect influences the dislocation core structure when the dislocation velocity approaches the limiting velocities, similar to Einstein's theory of relativity for particles approaching the speed of light \cite{relativistic1,eshelby1949uniformly}.
As the speed increases up to the Rayleigh wave speed, the interaction force between the screw components decreases while it increases for the edge components.
Hirth and Lothe determined the SFWs between partials in FCC and BCC metals via a Lagrangian formulation. They found that a screw dislocation core in FCC metals would constrict with increasing speed until it reaches $80\%$ the transverse sound speed where the trend is reversed and the dislocation core widens and eventually splits apart at the transverse sound speed \cite{hirth1968interactions}.
For dissociated edge dislocations approaching the Rayleigh wave speed, the attractive force between the opposite sign screw components will decrease and the repulsive force between the same signed edge components will increase \cite{weertman1962fast}.
This leads to the extension of the SFW as the speed approaches the Rayleigh wave and limiting speed. Once the dislocation's speed surpasses the limiting velocities, this relativistic effect is removed.
As a result, the dissociated edge dislocations constrict as seen in Figs. \ref{fig_edge_Al_Ni} and \ref{fig_edge_Au_Pt} for Ni and Au, respectively. For Au, the SFW expands again once the dislocation's velocity approaches the $2^\text{nd}$ limiting velocity.
To the authors' knowledge, this is the first validation of the relativistic effects on edge dislocation core structure that was previously proposed in theory \cite{weertman1962fast,hirth1968interactions}.

\subsection{Governing parameters for transonic dislocation motion}

The MD results in this work show certain trends for transonic dislocation motion behavior:

\begin{enumerate}
  \item Transonic motion is more frequently observed for edge dislocations than screw dislocations, which is consistent with previous MD simulation studies \cite{Olmsted_2005, Oren_2016, DANG2022}.
  \item Transonic motion is more frequently observed for FCC metals with large SFWs.
\end{enumerate}
These two results can be rationalized by understanding how transonic motion can be achieved. There are three intertwined factors that control this process: drag coefficient, instability, and limiting velocities, which depend on material properties such as stacking fault energies and elastic constants. 
For example, the drag coefficient is sensitive to the shear modulus \cite{BLASCHKE201924}.
The instability caused by nucleation of additional dislocations depends on the USF energy.
Finally, limiting velocities have been shown in previous studies \cite{Blaschke_2021} to be dependent on the elastic constants. However, it is most important to note that any unfavorable values of these parameters can prevent the transonic motion of dislocations.
For instance, a favorable drag coefficient and low limiting velocities can still fail to achieve dislocation transonic motion if the stress for instabilities to occur is too low (i.e. before the stress required for transonic motion can be reached).

Based on these three parameters, the first trend can be explained. For FCC metals, screw dislocations tend to have higher limiting velocities (see Table \ref{table:2}) but lower mobilities than edge dislocations.
This indicates a lower driving force and lower required target velocity for edge dislocations to achieve transonic motion compared to screw dislocations.
Note that while it is more challenging for screw dislocations to achieve transonic motion, it is still possible. For example, transonic screw dislocations have been observed in Cu \cite{Blaschke2021} at low temperature.
In this study, a low-temperature Au screw dislocation also approached the limiting velocity within $1\%$.

For the second trend, the effects of SFWs on the feasibility of dislocations to achieve transonic motion is clearly shown in this work. Specifically, Pt edge dislocations, with the smallest equilibrium SFW as shown in Table \ref{table:1}, do not reach transonic velocities. Similarly, Al edge dislocations, with the $2^\text{nd}$ smallest equilibrium SFW, only reach transonic velocity at low temperatures (namely 10K). On the other hand, Au and Ni edge dislocations with larger equilibrium SFWs both reach transonic velocities at both 10 and 100K. Small SFWs are equivalent to large ISF and USF energies as shown in previous studies \cite{hunter2011influence,DANG2019,Hunter:2014} where the dislocation core width is roughly inversely proportional to the product between these two stacking fault energies divided by the (averaged) shear modulus $G$.
With these large stacking fault energies, it is likely that the small SFWs in Pt dislocations (and Al dislocations to some extent) are less likely to expand when approaching the limiting velocities, which prevents them from crossing over to the transonic regime.
% Based on this, one can determine if a material system can achieve transonic dislocation velocity based on the SFW, ISF, and USF energies.
Overall, a material system's ability to achieve transonic dislocation velocity seems to depend on the interplay of the (slip system dependent) effective shear modulus (i.e. lowest limiting velocity), as well as the SFW, ISF, and USF energies.

\begin{figure}[ht]
\centering
\includegraphics[height=0.35\textwidth]{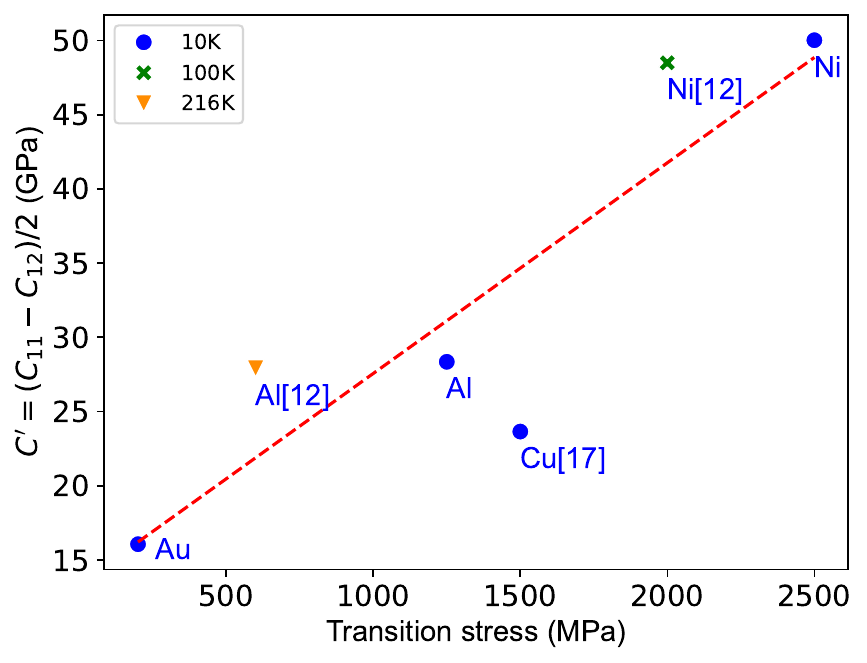}%
\includegraphics[height=0.35\textwidth]{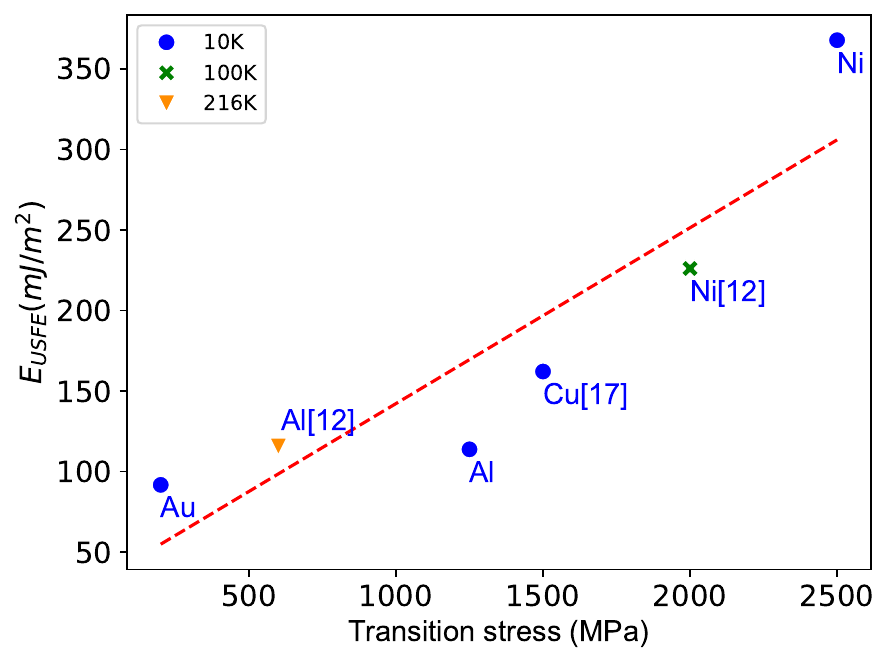}
\caption{$C'=\left(C_{11}-C_{12}\right)/2$ elastic constant (a) and unstable stacking fault (USF) energy (b) for different transition stresses of FCC metals in this study and previous MD simulation studies \cite{Olmsted_2005,Oren_2016}.
Elastic constants are given for 0K whereas the transition stress was determined at the temperatures indicated in the figure labels.}
\label{fig:Moduli}
\end{figure}

Predicting where and when the dislocations cross over to the transonic regime remains challenging. Figure \ref{fig:Moduli} shows the transition stresses, which are the lowest stress for the dislocations to move at velocities greater than a limiting velocity, at 100K and corresponding material properties/parameters. MD simulation data from \cite{Olmsted_2005} for Ni (100K) and Al (216K) as well as \cite{Oren_2016} for Cu (10K) are also included.
While the transition stress for FCC edge dislocations is generally proportional to both the $C'=\left(C_{11}-C_{12}\right)/2$ (which controls their lowest limiting velocity for the metals discussed here) and the USF energy (which is known to affect dislocation mobility, see \cite{Zhang:2023} and references therein), there was not a clear relationship between them.
One can argue that higher temperatures make it easier to reach transonic gliding velocities, so that the 216K and 100K data points (i.e. the lowest reported temperatures for Al and Ni in Ref. \cite{Olmsted_2005}) would move to the right if the temperature was reduced to 10K.
This would lead to better agreement with the linear trend of Fig. \ref{fig:Moduli}a and with our own simulation data at 10K (except for one outlier, Cu of Ref. \cite{Oren_2016}).
Attempts to connect to other parameters also provide no clear trends with the transition stresses. This is to be expected since the transonic dislocation motion relies on the interplay between three parameters as mentioned earlier.

%\clearpage
\section{Conclusions}

In conclusion, the transonic motion of dislocations in FCC metals are systematically studied using MD simulations of edge and screw dislocations.
It was found that transonic velocities are easier to achieve by edge dislocations compared to screw dislocations due to their higher mobilities and lower limiting velocities.
Moreover, the dislocation core is found to widen as the edge dislocation approaches the limiting velocities and constrict again once the dislocation reaches the transonic velocity, which is due to the relativistic effects.
Importantly, it is also found that the dislocation's ability to reach transonic velocities is proportional to its SFW.
While this study only focuses on FCC metals, these results highlight the importance of the dislocation core structure to dislocation mobility in the extreme regime.

\subsection*{Acknowledgements}

We thank the two reviewers for numerous helpful comments.
This work was supported by the Advanced Technology Development and Mitigation (ATDM) project and the Physics and Engineering Models (PEM) Materials project within the Advanced Simulation and Computing (ASC) Program of the U.S. Department of Energy under contract 89233218CNA000001.
This research
used resources provided by the Los Alamos National Laboratory Institutional Computing Program, which is supported
by the U.S. Department of Energy National Nuclear Security
Administration under Contract No. 89233218CNA000001.

\bibliographystyle{IEEETran_custom}
\bibliography{sample}

\end{document}